\documentclass[showpacs,preprintnumbers,amsmath,amssymb,twocolumn]{revtex4}
  \usepackage{amsmath}
\usepackage{graphicx}
\usepackage{epsf}
\usepackage{psfrag}
\usepackage{epsfig}
\usepackage{graphics}

\newcommand{\be}{\begin{equation}}
\newcommand{\ee}{\end{equation}}
\newcommand{\bq}{\begin{eqnarray}}
\newcommand{\eq}{\end{eqnarray}}

\begin{document}

\title{{Ambiguities in the gravitational correction of quantum electrodynamics}}

\author{J. C. C. Felipe$^{(a)}$}\email[]{guaxu@fisica.ufmg.br}
\author{L. A. Cabral$^{(a,c)}$}\email[]{cabral@uft.edu.br}
\author{L. C. T. Brito$^{(b)}$} \email []{lcbrito@dex.ufla.br}
\author{Marcos Sampaio$^{(a)}$} \email []{msampaio@fisica.ufmg.br}
\author{M. C. Nemes$^{(a)}$}\email[]{mcnemes@fisica.ufmg.br}

\affiliation{(a) Universidade Federal de Minas Gerais - Departamento de F\'{\i}sica - ICEX \\ P.O. BOX 702,
30.161-970, Belo Horizonte MG - Brazil}
\affiliation{(b) Universidade Federal de Lavras - Departamento de Ci\^encias Exatas\\
P. O. BOX 3037, 37.200-000, Lavras MG - Brazil}
\affiliation{(c) Universidade Federal do Tocantins - Centro de Ci\^{e}ncias\\
P. O. BOX 132, 77.804-970, Aragua\'{\i}na TO - Brazil}

\begin{abstract}
We verify that quadratic divergences stemming from gravitational corrections to QED which have been conjectured to lead to asymptotic
freedom near Planck scale are arbitrary (regularization dependent) and compatible with zero. Moreover we explicitly show that such arbitrary
term contributes to the beta function of QED in a gauge dependent way in the gravitational sector.
\noindent
\end{abstract}
\maketitle

Gravitational effects in Quantum Field Theory can only be obtained in the context of effective theories. This situation resembles the early
days of Quantum Mechanics when the study of matter-radiation interaction was treated in a similar fashion. Matter was quantized while
electromagnetic field was treated classically. Nonetheless in both cases relevant information has been extracted from such effective
theories. In the case of quantum gravity the issue of effective theories has been discussed in depth in
\cite{'tHooft:1974bx,Deser:1974cz,Deser1974cs,Weinberg:1978kz,Donoghue:1993eb}.

Recently a most challenging far reaching result has been put forth by Robinson and Wilczek. They showed that
quantum gravitational effects may contribute to asymptotic freedom in gauge theories \cite{Robinson:2005fj}. As to be expected several independent investigations
started to appear arguing either in favor or against the result. Within the context of the Einstein-Yang-Mills
model, in \cite{Ebert:2007gf} it is argued that the quadratic divergences ultimately responsible for asymptotic freedom are canceled
in some step of the calculation. In \cite{Folkerts2012} the gauge dependence of the gluon and graviton propagator are included and
asymptotic freedom remains, being however gauge dependent in what concerns the gravitational sector. Results in the context of the Einstein-
Maxwell model, are also contradictory. The issue of the gauge dependence has been addressed in ref \cite{Pietrykowski:2006xy}. Also in
\cite{Ebert:2007gf,Toms:2010vy,TangWu2012} the quadratic divergences appear again as the agent of asymptotic freedom in this model. In
\cite{Felipe2011} it is shown that quadratic divergences are intrinsically ambiguous in a regularization independent way. Other regularization
dependence may arise during the calculations since surface terms will certainly appear if quadratic divergences are present and are evaluated
differently in different regularizations. The authors of \cite{Ellis:2012} also argue that quadratic divergences are devoid of physical
significance, since they may be eliminated by a redefinition of the fields in the original Lagrangian. An argument along a different line was given
recently in ref \cite{Nielsen:2011} where it is argued that the linearized spacetime metric is not a solution of Einstein equation. Against
the existence of asymptotic freedom in the context of the Einstein-Maxwell model a calculation is presented which leaves arbitrary gauge
parameters (both for the photon and graviton sectors) and concludes that the results are dependent on the gravitational gauge only \cite{Leonard:2012}.
Moreover it is shown that in the limit of zero mass quadratic divergences disappear, in agreement with Dimensional Regularization. 

Given the nature of the discrepancies in the results in the literature regarding  both gauge and regularization dependence we believe
it is pertinent to present a calculation using the most general parametrization of logarithmic and quadratic integrals which allows one
to display ambiguities arising either from divergences as well as those intrinsic to perturbative calculations such as surface terms. We
carry all ambiguities till the end of the calculation. 

We show that there appears in fact a quadratic divergence which could in principle be responsible for asymptotic freedom. It is
however intrinsically ambiguous as it is both regularization and gauge (in the gravitational sector) dependent.

We start by describing the model. We couple scalar electrodynamics with gravitation as follows
\begin{eqnarray}\label{actiontotal}
  S&=&\int d^{4}x \sqrt{-g} \Bigg[ -\frac{1}{4}(g^{\mu\alpha}g^{\nu\beta}F_{\mu\alpha}F_{\nu\beta}) + \nonumber \\&& 
g^{\mu\nu}(D_{\mu}\phi)(D_{\nu}\phi)^{*} - m^{2}{\phi}{\phi}^{*} + \frac{2}{\kappa^{2}}  R \Bigg],
\end{eqnarray}
where in (\ref{actiontotal}),  $F_{\mu\nu}$ is the (Maxwell) field strength
tensor, $(D_{\mu}\phi)=(\partial_{\mu}\phi+ieA_{\mu}\phi)$ represents the covariant derivative, $g_{\mu\nu}$ the space time metric, $\kappa=\sqrt{32\pi G}$ the gravitational coupling constant and $R$ the scalar curvature
tensor. We use the weak field approximation, 
linearizing the metric around a Minkowski background $\eta_{\mu\nu}=(1,-1,-1,-1)$ in the following way
\begin{equation}\label{weakfield}
 g_{\mu\nu}=\eta_{\mu\nu}+\kappa h_{\mu\nu},
\end{equation}
with $h_{\mu\nu}$ in eq. (\ref{weakfield}) being the deviation from the flat background.

The
 propagator of the scalar field is 
\begin{equation}\label{scalarprop}
{\Delta}(p)=\frac{i}{p^{2}-m^{2}},
\end{equation}
whereas the graviton propagator with explicit gauge dependence reads
\begin{equation}\label{gravitprop}
\Delta^{\alpha\lambda\sigma\beta}(k)=\frac{iP^{\alpha\lambda\sigma\beta}}{(k^{2}-\mu^{2})}+\frac{1}{2}({\xi}-1) \; \frac{iM^{\alpha\lambda\sigma\beta}}{(k^{2}-\mu^{2})^{2}}, 
\end{equation}
with
\begin{equation}\label{numerator}
P^{\alpha\lambda\sigma\beta}=\frac{1}{2}\Big(\eta^{\beta\lambda}\eta^{\sigma\alpha} + \eta^{\beta\alpha}\eta^{\lambda\sigma} - \eta^{\alpha\lambda}\eta^{\sigma\beta}\Big),
\end{equation}
\begin{eqnarray}
M^{\alpha\lambda\sigma\beta}&=&\big(\eta^{\alpha\sigma}(k^{\lambda}k^{\beta})+\eta^{\alpha\beta}(k^{\lambda}k^{\sigma})+\eta^{\lambda\sigma}(k^{\alpha}k^{\beta})\nonumber \\&&
+\eta^{\lambda\beta}(k^{\alpha}k^{\sigma})\big)
\end{eqnarray}
and $\mu$ is a fictitious mass which we will set to zero at the end of calculation. The photon propagator as usual is given by
\begin{equation}\label{propphoton}
\Delta^{\mu\nu}(k)=\frac{-i\eta^{\mu\nu}}{(k^{2}-\mu^{2})}- i\frac{(1-\alpha)}{\alpha}\frac{k^{\mu}k^{\nu}}{(k^{2}-\mu^{2})^{2}},
\end{equation}
in an arbitrary gauge $\alpha$.

\begin{figure}[!h]
\begin{minipage}[!h]{0.99\linewidth}
\includegraphics[width=\linewidth]{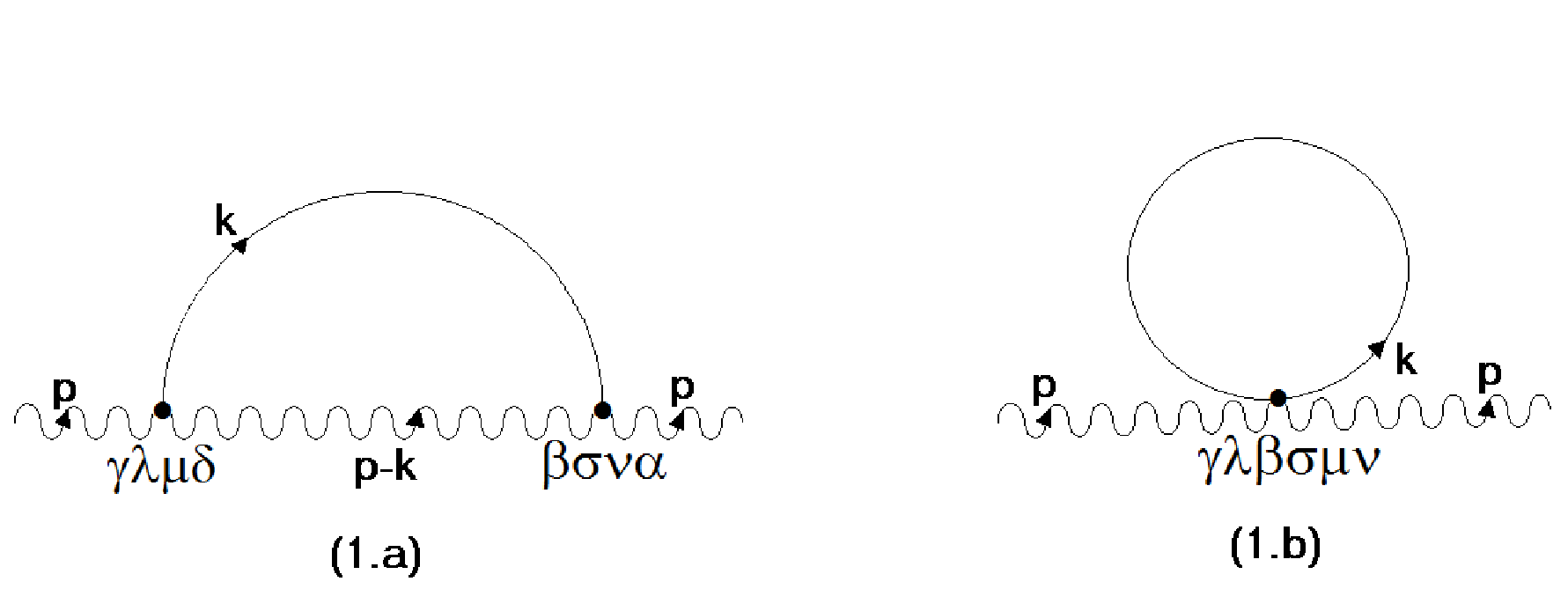}
\caption{One-loop gravitational corrections to the photon propagator. Solid and wavy lines represent graviton and photon propagators respectively.}
\label{gravity}
\end{minipage} \hfill
\end{figure}

\begin{figure}[!h]
\begin{minipage}[!h]{0.99\linewidth}
\includegraphics[width=\linewidth]{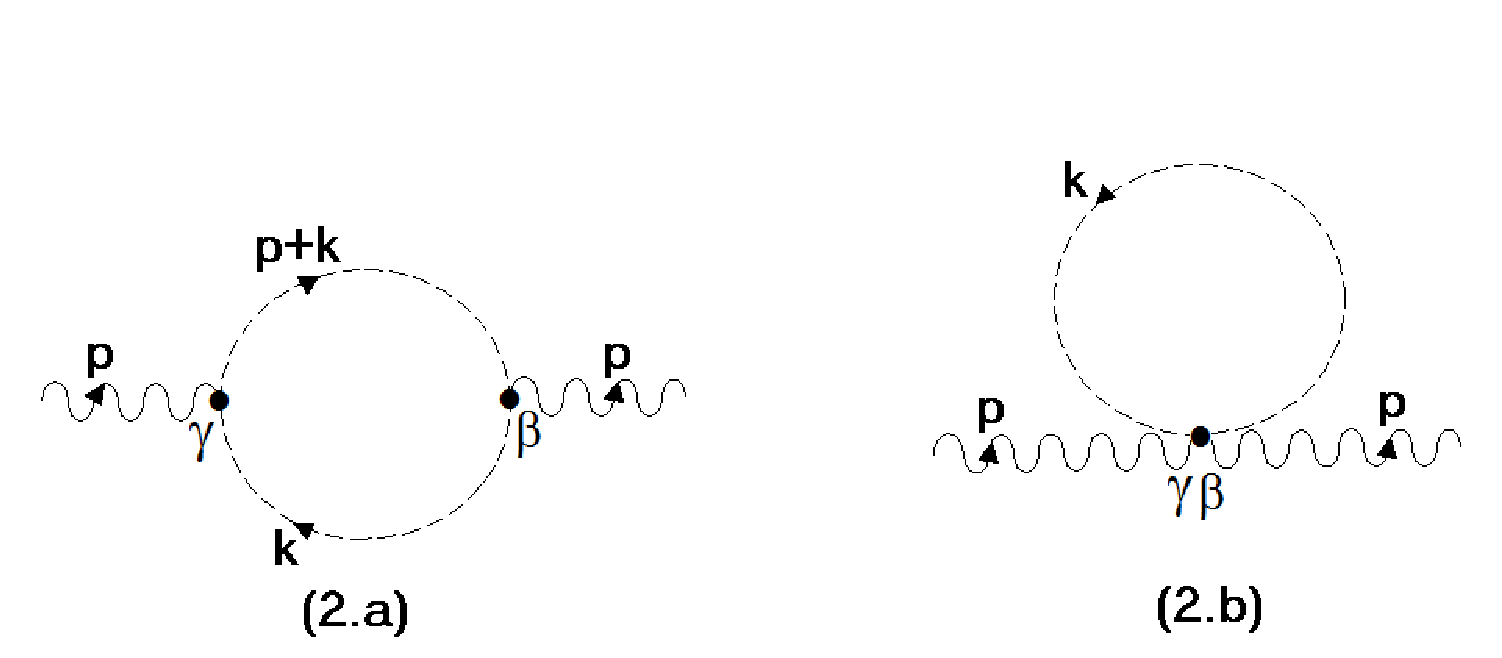}
\caption{One-loop matter field corrections to the photon propagator. Dashed line represents the scalar field.}
\label{scalar}
\end{minipage} \hfill
\end{figure}

The one loop corrections to the photon propagator for scalar QED in weak gravitational field approximation (\ref{weakfield}) are depicted in figures (\ref{gravity}) and (\ref{scalar})
respectively.

The graviton-photon vertices in fig. (\ref{gravity}) are given by

\begin{eqnarray}\label{verticgravity}
& & V^{\lambda\theta\gamma\delta}(p,p^{\prime}) =  i\kappa\Big\{P^{\lambda\theta\gamma\delta}(p\cdot p') \nonumber\\
& &+\frac{1}{2}\left[ \eta^{\gamma\delta}\left(p^{\theta}p'^{\lambda}+p^{\lambda}p'^{\theta}\right) \right.
+\eta^{\lambda\theta}p^{\delta}p'^{\gamma}
- \eta^{\lambda\delta}p'^{\gamma}p^{\theta}\nonumber\\
& &-\eta^{\lambda\gamma}p^{\delta}p'^{\theta}
-\left.\eta^{\theta\gamma}p^{\delta}p'^{\lambda}-\eta^{\theta\delta}p^{\lambda}p'^{\gamma}\right]\Big\},
\end{eqnarray}
where, the term $P^{\lambda\theta\gamma\delta}$ is given by in (\ref{numerator}).
\begin{eqnarray}\label{gravitonvertex}
& &V^{\sigma\tau\theta\delta\rho\xi}(p,p')=i\frac{\kappa^{2}}{4}\Big\{\eta^{\delta\theta}\eta^{\rho\sigma}\tilde{\Omega}^{\tau\xi} +2\eta^{\tau\theta}\eta^{\sigma\delta}\tilde{\Omega}^{\rho\xi}\nonumber \\
& &+\eta^{\sigma\tau}\Big\{\eta^{\xi[\rho}\Omega^{\delta]\theta}+\eta^{\xi[\rho} \Omega^{\theta]\delta}-\eta^{\rho(\delta} \Omega^{\theta)\xi}\Big\}\nonumber \\
& &+\eta^{\delta\theta}\Big\{\eta^{\xi[\rho} \Omega^{\sigma]\tau}+\eta^{\xi[\rho} \Omega^{\tau]\sigma}+\eta^{\tau[\sigma} \Omega^{\rho]\xi}\Big\}\nonumber \\
& &+2\Big\{\eta^{\xi[\sigma}\Omega^{\delta][\tau}\eta^{\theta]\rho}+\eta^{\xi[\tau}\Omega^{\theta][\sigma}\eta^{\delta]\rho}\Big\}\nonumber \\
& &+4\eta^{\sigma\delta}\Big\{\eta^{\rho(\theta} \Omega^{\tau)\xi}+\eta^{\xi[\theta} \Omega^{\rho]\tau}+\eta^{\xi[\tau} \Omega^{\rho]\theta}\Big\}\nonumber \\
& &+(p \cdot p')\Big\{\eta^{\delta\theta}\eta^{\tau[\rho} \eta^{\sigma]\xi}+\eta^{\tau\sigma}\eta^{\theta(\rho} \eta^{\delta)\xi}\nonumber \\
& &-4\eta^{\sigma\delta}\eta^{\theta(\rho} \eta^{\xi)\tau}\Big\}\Big\},
\end{eqnarray}
where $\Omega^{\mu\nu}=p^{\mu}p'^{\nu}$, $\tilde{\Omega}^{\mu\nu}=\eta^{\mu\nu}(p \cdot p')-{\Omega}^{\mu\nu}$ 
and the parentheses (brackets) indicate symmetrization (anti-symmetrization) in the indices. In fig. (\ref{scalar}), the photon scalar field vertices are
\begin{equation}\label{verticescalar}
V^{\gamma}(p,p^{\prime})= -ie(p+p^{\prime})^{\gamma}
\end{equation}
and
\begin{equation}\label{verticescalar1}
V^{\gamma\beta}= 2ie^{2}\eta^{\gamma\beta}
\end{equation}
respectively.

The photon self energy correction to one loop order is given by the sum of the amplitudes depicted in figs. (\ref{gravity}) and
(\ref{scalar}). They read 
\begin{eqnarray}\label{one_loop}
\Pi^{\mu\nu}_{1a}(p)&=&\int_k
 \Delta_{\alpha\delta}(k)V^{\gamma\lambda\mu\delta}(p,p-k)\Delta_{\gamma\lambda\beta\sigma}(k)\nonumber \\
& &\times V^{\beta\sigma\nu\alpha}(p-k,p),
\end{eqnarray}
\begin{equation}\label{one_loop_1}
\Pi^{\mu\nu}_{1b}(p)=\int_k V^{\alpha\lambda\sigma\beta\mu\nu}(p,p')\Delta_{\alpha\lambda\sigma\beta}(k),
\end{equation}
\begin{eqnarray}\label{oneloopscalar}
\Pi^{\mu\nu}_{2a}(p)&=&
 \int_k  V^{\mu}(p+2k)\Delta(p)V^{\nu}(p+2k)\nonumber \\
& & \times \Delta(p+k),
\end{eqnarray}
\begin{equation}\label{oneloopscalar_1}
 \Pi^{\mu\nu}_{2b}(p)=\int_k V^{\mu\nu} \Delta(p),
\end{equation}
where, for brevity, $\int_k=\int \frac{d^{4}k}{(2\pi)^{4}}$.

To evaluate these amplitudes we adopt a strategy based on Implicit Regularization \cite{IR,IR2}.
Their ultraviolet content will be displayed in terms of basic divergent integrals which depend on the loop momentum only.
The difference between basic divergent integrals which have the same superficial degree of divergence but different Lorentz structure in
the internal momenta can be cast as surface terms which are in principle arbitrarily valued. In \cite{Luellerson2012}
it was shown that such arbitrariness can be fixed on symmetry grounds (e.g. gauge and supersymmetry) which is ultimately related to 
momentum routing invariance in the Feynman diagrams.
Typical basic divergent integrals at one loop order are given by 
\begin{equation}\label{divergence_log}
I_{log}(\mu^{2})=\int_{k} \frac{1}{(k^{2}-\mu^{2})^{2}} 
\end{equation}
and
\begin{equation}\label{divergence_quad}
I_{quad}(\mu^{2})=\int_{k}\frac{1}{(k^{2}-\mu^{2})}.
\end{equation}

Other basic loop divergent integrals appear in the calculation, e.g.
\begin{equation}\label{logmunu}
I_{log}^{\mu\nu}(\mu^{2})=\int_k \frac{k^{\mu}k^{\nu}}{(k^{2}-\mu^{2})^{3}}
\end{equation}
and
\begin{equation}\label{quadmunu}
I_{quad}^{\mu\nu}(\mu^{2})=\int_k \frac{k^{\mu}k^{\nu}}{(k^{2}-\mu^{2})^{2}},
\end{equation}
which however are related to (\ref{divergence_log}) and (\ref{divergence_quad}) through surface terms.

We can parametrize the divergent integrals in a general form \cite{Luellerson2012}. For example, $I_{log}(\mu^2)$ in equation (\ref{divergence_log}) satisfies the regularization independent
equation
\begin{equation}\label{dilog}
\frac{dI_{log}(\mu^{2})}{d\mu^{2}}=-\frac{b}{\mu^{2}},
\end{equation}
where $b=-\frac{i}{(4\pi^{2})}$. The most parametrization compatible with equation (\ref{dilog}) is
\begin{equation}\label{dilog_p}
I_{log}(\mu^{2})= b_{1}+b\ln\Bigg(\frac{\Lambda^{2}}{\mu^{2}}\Bigg),
\end{equation}
with $b_{1}$ is arbitrary constant. In analogous fashion, $I_{log}^{\mu\nu}(\mu^{2})$ can be written
\begin{equation}\label{dilog_munu}
I_{log}^{\mu\nu}(\mu^{2})= \frac{g^{\mu\nu}}{4}\Bigg[b_{1}'+bln\Bigg(\frac{\Lambda^{2}}{\mu^{2}}\Bigg)\Bigg],
\end{equation}
where $b'_{1}$ is another arbitrary constant. The choice of the arbitrary constant have a physical motivation. Notice that
\begin{eqnarray}\label{consistentrelation}
g^{\mu\nu}I_{log}(\mu^{2})-4I^{\mu\nu}_{log}(\mu^{2})&=&\int_{k} \frac{\partial}{\partial k_{\mu}}\frac{k^{\nu}}{(k^{2}-\mu^{2})^{2}}\nonumber \\
&=&g^{\mu\nu}(b_{1}-b'_{1})
\end{eqnarray}
is a surface term, which is in principle arbitrary and generally regularization dependent as the parametrization shown in eqs. (\ref{dilog_p}) and
(\ref{dilog_munu}).

Likewise, equation (\ref{divergence_quad}) satisfies
\begin{equation}\label{diquad_p}
\frac{dI_{Iquad}(\mu^{2})}{d\mu^{2}}=I_{log}(\mu^{2})
\end{equation}
and the most consistent parametrization from (\ref{diquad_p}) is
\begin{equation}\label{diquad}
I_{quad}(\mu^{2})= c_{1}\Lambda^{2} + b\mu^{2}\ln\Bigg(\frac{\Lambda^{2}}{\mu^{2}}\Bigg) +c'_{1}\mu^{2}.
\end{equation}
and, in analogous way
\begin{equation}\label{diquad_munu}
I^{\mu\nu}_{quad}(\mu^{2})=\frac{g^{\mu\nu}}{2}\Bigg[c_{2}\Lambda^{2} + b\mu^{2}\ln\Bigg(\frac{\Lambda^{2}}{\mu^{2}}\Bigg)+c'_{2}\mu^{2}\Bigg],
\end{equation}
These parametrizations reveal the regularization dependent character of quadratic divergence expressed by
the dimensionless constants $c_{i}$ and $c^{\prime}_{i}$. For instance, whilst $c_{1}=c_{2}=0$ in dimensional regularization,
in sharp momentum cutoff they evaluate to
\begin{equation}\label{cutoffquadterm}
c_{1}=c_{2}=-\frac{i}{64\pi^{2}}.
\end{equation}
In other words, while gauge invariance and momentum routing fix the surface terms to zero, $c_{1}$, for instance, remains undetermined.

After this brief discussion about regularization ambiguities, we proceed to present our main results. 

A lengthy yet straightforward calculation we get
\begin{eqnarray}\label{gravpolatensor} 
& &{\Pi}^{\mu\nu}_{total}(p)=-\kappa^{2}\bigg[\frac{(9{\xi}+1)}{24} F(p^{2})p^{2}\nonumber \\
& &+\frac{(13{\xi}-5)}{4}I_{quad}(\mu^{2})+\frac{(1-\alpha)}{6\alpha}F(p^{2})p^{2}\nonumber \\
& &+\frac{(\xi-1)(1-\alpha)}{4}F(p^{2})p^{2}\Bigg]\big(p^{2}\eta^{\mu\nu}-p^{\mu}p^{\nu}\big)\nonumber \\
&&-\kappa^{2}\Bigg[\Upsilon^{\mu\nu}_{2}
+\frac{(1-\alpha)}{\alpha}\Upsilon^{\mu\nu}_{3}+({\xi}-1)\Upsilon^{\mu\nu}_{4}\nonumber \\
& &+({\xi}-1)\frac{(1-\alpha)}{\alpha}\Upsilon^{\mu\nu}_{5}\bigg]\nonumber \\
& &-e^{2}\bigg[\frac{1}{3}F(p^{2})(\eta^{\mu\nu}p^{2}-p^{\mu}p^{\nu})-4\Upsilon^{\mu\nu}_{1}\bigg],
\end{eqnarray}
where the term $F(p^2)$ in equation (\ref{gravpolatensor}) is given by
\begin{equation}
F(p^2)=I_{log}(\mu^2)- \frac{i}{16 \pi^2} \ln \Bigg( -\frac{p^2}{\mu^2} \Bigg)
\end{equation}

The infrared regulator expressed by $\ln(\mu^{2})$ when $\mu\rightarrow 0$ in finite part of the amplitude is canceled against the infrared
regulator in $I_{log}(\mu^{2})$ through the regularization independent relation
\begin{equation}\label{scalarrelation}
I_{log}(\mu^{2})=I_{log}(\lambda^{2})-\frac{i}{16\pi^{2}}\ln\Bigg(\frac{\mu^{2}}{\lambda^{2}}\Bigg),
\end{equation}
where $\lambda^2$ is a non vanishing arbitrary parameter which plays the role of renormalization group constant \cite{Brito:2008zn}.

In (\ref{gravpolatensor}) the (regularization dependent) surface terms read

\begin{equation}\label{surface1}
\Upsilon_{1}^{\mu\nu}={a_{2}}\eta^{\mu\nu} - \Bigg(a_{1}-\frac{1}{6}a_{3}\Bigg)(\eta^{\mu\nu}p^{2} + 2p^{\mu}p^{\nu}),
\end{equation}
\begin{eqnarray}\label{surface2}
& &\Upsilon_{2}^{\mu\nu}=\frac{a_{1}}{12}p^{2}\left(13p^{2}\eta^{\mu\nu}-20p^{\nu}p^{\mu}\right) \nonumber \\
& &- \left[\frac{a_{2}}{2} - \frac{8 a_{3}}{3}\, p^{2} \right]\left(\eta^{\mu\nu}p^{2}-p^{\mu}p^{\nu}\right),
\end{eqnarray}
\begin{equation}\label{surface3}
\Upsilon_{3}^{\mu\nu}=(\eta^{\mu\nu}p^{4}-p^{2}p^{\mu}p^{\nu})[2a_{3}-a_{1}],
\end{equation}
\begin{equation}
\Upsilon^{\mu\nu}_{4}=\Bigg[2(a_{3}+a_{1})p^{2}+a_{2}\Bigg]
(p^{2}\eta^{\mu\nu}-p^{\mu}p^{\nu}\big),
\end{equation}
\begin{equation}
\Upsilon^{\mu\nu}_{5}= \Bigg[\frac{3a_{1}}{2}-a_{3}\Bigg]p^{2}\Big(p^{2}\eta^{\mu\nu}-p^{\mu}p^{\nu}\Big),
\end{equation}
where the terms $a_{1}$, $a_{2}$ and $a_{3}$ are given by
\begin{equation}\label{c_1}
a_{1}\eta^{\mu\nu}=\frac{1}{4}\eta^{\mu\nu}I_{log}(\mu^{2})-\int_k\frac{k^{\mu}k^{\nu}}{(k^{2}-\mu^{2})^{3}},
\end{equation}
\begin{equation}\label{c_2}
a_{2}\eta^{\mu\nu}=\frac{1}{2}\eta^{\mu\nu}I_{quad}(\mu^{2})-\int_k\frac{k^{\mu}k^{\nu}}{(k^{2}-\mu^{2})^{2}},
\end{equation}
and
\begin{eqnarray}\label{c_3}
& &a_{3}\eta^{\{ \mu\nu} \eta^{\alpha\beta\}}=\frac{1}{24}\eta^{\{ \mu\nu} \eta^{\alpha\beta\}}I_{log}(\mu^{2})\nonumber \\
& &-\int_k \frac{k^{\mu}k^{\nu}k^{\alpha}k^{\beta}}{(k^{2}-\mu^{2})^{4}},
\end{eqnarray}
respectively. 

Whilst quadratic divergences are canceled between the graphs in figure (2) there remains
a $I_{quad}(\mu^2)$ from the graphs in figure (1). This term proportional to $I_{quad}(\mu^2)$ in
(\ref{gravpolatensor}) could give rise to asymptotic freedom 
in electrodynamics as pointed out by Toms in \cite{Toms:2010vy}. We show however that besides being gauge 
dependent it is also intrinsically undetermined as we can see in eqs (\ref{diquad}) and (\ref{diquad_munu}) respectively. As for the
other indeterminacies contained in eq. (\ref{gravpolatensor}) expressed by the $\Upsilon's$ (surface terms) it can be easily
checked that the terms $a_i$, with $i=1,...,3$ are evaluated to zero in dimensional regularization, rendering $\Pi^{\mu\nu}_{total}$
transverse, as it should. Within the framework of Implicit Regularization we have demonstrated that both
momentum routing and gauge invariance are simultaneously satisfied if we set surface terms to zero \cite{Luellerson2012}. Results in the
literature can be recovered by choosing $\xi=\alpha=1$.

We get the $\beta$ function from eq. (\ref{gravpolatensor}). It is given by
\begin{equation}\label{betafunction}
\beta= \frac{e^{3}}{12\pi^{2}}-(13{\xi}-5)\frac{ic_{1}}{2}{\Lambda}^{2} \kappa^2 e,
\end{equation}
where $c_{1}$ in eq. (\ref{betafunction}) remains arbitrary and can not be fixed by gauge invariance. In the same sense, \cite{MohamedDonoghue2011,Toms2012,MohamedDonoghue2012} also argue that the quadratic divergence present in (\ref{gravpolatensor}) should not exist because
the quadratic divergence dependence disappears when physical processes are taking into account.

To summarize we have shown in a regularization independent way that the quadratic divergences which contribute to asymptotic freedom
in QED at the Planck scale is intrinsically undetermined and gauge dependent.

A final comment is in order. In references \cite{MohamedDonoghue2011,Toms2012,MohamedDonoghue2012} it is argued that the cutoff dependence
represented by the cutoff $\Lambda$ should not affect physical observables while in \cite{Harst2011} and \cite{Daum2010} it is performed an Asymptotic Safe construction
of QED coupled to Quantum Einstein Gravity attaching physical meaning to the cutoff. Because the quadratic divergence ambiguity expressed by
$c_{1}$ is arbitrary and compatible with positive values, it reveals an essential ambiguity in gravitational corrections to the running
charge and thus cannot affect physical observables.

Acknowledgements: This work was supported by CAPES, CNPq and FAPEMIG.


\begin{thebibliography}{}
 \bibitem{'tHooft:1974bx}
  G.~'t Hooft and M.~J.~G.~Veltman,
  Annales Poincare Phys.\ Theor.\  A {\bf 20}, 69 (1974).

\bibitem{Deser:1974cz}
  S. Deser and P. Van Nieuwenhuizen,
  Phys.\ Rev.\  D {\bf 10}, 401 (1974).

\bibitem{Deser1974cs}
S Deser and P. Van Nieuwenhuizen,
Phy. \ Rev. \ Lett. {\bf 32}, 245-247 (1974).

\bibitem{Weinberg:1978kz}
  S.~Weinberg,
  Physica A {\bf 96}, 327 (1979).

\bibitem{Donoghue:1993eb}
  J.~F.~Donoghue,
  Phys.\ Rev.\ Lett.\  {\bf 72}, 2996 (1994).

\bibitem{Robinson:2005fj}
  S.~P.~Robinson and F.~Wilczek,
  Phys.\ Rev.\ Lett.\  {\bf 96}, 231601 (2006).

\bibitem{Ebert:2007gf}
  D.~Ebert, J.~Plefka and A.~Rodigast,
  Phys.\ Lett.\  B {\bf 660}, 579 (2008).

\bibitem{Folkerts2012}
  Sarah Folkerts, Daniel F. Litim, Jan M. Pawlowski,
  Phys. \ Lett. \ B {\bf 709}, 234-241 (2012).

\bibitem{Pietrykowski:2006xy}
  A.~R.~Pietrykowski,
  Phys.\ Rev.\ Lett.\  {\bf 98}, 061801 (2007).


\bibitem{Toms:2010vy}
  D.~J.~Toms,
  Nature {\bf 468}, 56 (2010).

\bibitem{TangWu2012}
  Yong Tang and Yue-Lian Wu,
  Commu. \ Theor. \ Phys. {\bf 57}, 629-636 (2012).

\bibitem{Felipe2011}
  J.~C.~C.~Felipe, L.~C.~T.~Brito, M.~Sampaio, M.~C.~Nemes,
  Phys.\ Lett.\  B {\bf 700}, 86-89 (2011).

\bibitem{Ellis:2012}
J.~Ellis, N.~E.~Mavromatos, 
  Phys. \ Lett. \ B {\bf 711}, 139 (2012).

\bibitem{Nielsen:2011}
  N. K. Nielsen,
  Annals \ Phys. {\bf 327} 861-892 (2012).

\bibitem{Leonard:2012}
   K. E. Leonard and R. P. Woodard,
   arXiv: 1202.5800v1. (2012).

\bibitem{IR}
A. P. B. Scarpelli, M. Sampaio and M. C. Nemes, \ Phys.  Rev. \ D \textbf{63} (2001) 046004.

\bibitem {IR2}
A. P. B. Scarpelli, M. Sampaio, M. C. Nemes and B. Hiller, \ Phys. Rev. \ D \textbf{64} (2001) 046013.

\bibitem{Luellerson2012}
  L. C. Ferreira, A. L. Cherchiglia, B. Hiller, M. Sampaio, M. C. Nemes
  arXiv: 1110.6186, to appear in Phys. \ Rev. \ D (2012).

\bibitem{Brito:2008zn} 
  L.~C.~T.~Brito, H.~G.~Fargnoli, A.~P.~B. Scarpelli, M.~Sampaio and M.~C.~Nemes,
  Phys.\ Lett.\  B {\bf 673}, 220 (2009) and references therein.

\bibitem{MohamedDonoghue2011}
  M. M. Anber, J. F. Donoghue and M. El-Houssieny,
  Phys. \ Rev. \ D {\bf 83} (2011) 124003.

\bibitem{Toms2012}
  D. J. Toms,
  Phys. \ Rev. \ D {\bf 84} (2011) 084016.

\bibitem{MohamedDonoghue2012}
  M. M. Anber and J. F. Donoghue,
  Phys. \ Rev. \ D {\bf 85} (2012) 104016.

\bibitem{Harst2011}
  U. Harst and M. Reuter,
  JHEP {\bf 05} (2011) 119.

\bibitem{Daum2010}
  J.-E. Daum, U. Harst and M. Reuter,
  JHEP {\bf 01} (2010) 84.
  
  

\end{thebibliography}
\end{document}